\documentclass[11pt,reqno]{amsart}
\usepackage{t1enc}
\usepackage[latin1]{inputenc}
\usepackage{amsmath,latexsym,amssymb,graphicx,dsfont,amsthm,amsfonts,eurosym}
\usepackage{color}
\usepackage{tikz}
\usepackage{hyperref}
\usetikzlibrary{arrows.meta}
\usepackage{yfonts}
\usepackage{float,tabls}
\usepackage{umoline,comment}

\numberwithin{equation}{section}

\makeatletter
\@namedef{subjclassname@2020}{%
  \textup{2020} Mathematics Subject Classification}
\makeatother

\begin{document}

\title[Smoking Cessation via Online Hypnosis]{A Data-Analytic Evaluation of Smoking Cessation via Online Hypnosis}

\author{Lorenz A. Gilch}
\address{Lorenz A. Gilch: University of Passau, Innstr. 33, 94032 Passau, Germany}

\email{Lorenz.Gilch@uni-passau.de}
\urladdr{http://www.math.tugraz.at/$\sim$gilch/}

\author{Alexandra Wojak}
\address{Alexandra Wojak: Mittlere M\"uhle 22, 45665 Recklinghausen, Germany}

\email{Alexandra.Wojak@gmail.com}
\urladdr{http://www.praxis-atempause-online.de}


\date{\today}

\maketitle



\begin{abstract}
In this article we report the observations of our field study which investigates the success rates of online hypnosis for smoking cessation. Following an international call in Germany, Austria and Switzerland  for subjects willing to stop smoking  $83$ participants contributed to this study, which took place during the lockdowns of the COVID-19 pandemic. All subjects received two online hypnosis sessions in small groups and additionally got two MP3 audio files for invididual intensification. The participants sent back several questionnaires, from which we present the evaluation results, and we discuss the problems observed during the study. It turned out that $86\%$ of the participants are smoke-free after $6$ months as long as they participated according to the intervention protocol. Among all smokers after six months we have observed a reduction of $32\%$ in cigarette consumption. This report shall be seen as a statistical evaluation of the collected data and not as the report of a fully pre-designed study in a narrower sense.
\end{abstract}

\section{Introduction}
\label{sec:introduction} 

According to the German Federal Ministry of Health \cite{bmg-tote} smoking is the most avoidable health risk in Germany: each year in average $127.000$ deaths in Germany are counted which are related to tobacco consumption. In 2019 about $458.000$ patients have received hospital treatments due to smoking-specific diseases like lung cancer or COPD, which is an increase of $18\%$ compared to the year 2010, see German Federal Office of Statistics \cite{destatis1}. Governmental programs have been established in order to decrease the rate of smokers. Morever,
during the last decade anti-smoking laws in Germany and across Europe have been tightened in order to protect non-smokers in public areas like train stations, airports or the  interior of restaurants. This seems to imply an increasing demand of smokers in supporting them to quit smoking, compare with German Federal Ministry of Health \cite{bmg-rauchfrei}.
\par
Many smokers seem to think about quitting smoking, but most of them are not successful without any additional help or intervention and there are high fall-back rates after some time without smoking. According to Hughes, Keely and Naud \cite{hughes-keely-naud} and Giovino et al. \cite{giovino} only $3\%-5\%$ of smokers, who try to quit on their own, are successful after $6$ to $12$ months.
There is a big variety of methods to support smokers in their wish to become abstinent of cigarettes, e.g., single and group consultations, pharmalogical interventions, behaviour and/or psychological therapies and also intervention by hypnosis. In Germany the federal government proposed the program \textit{Rauchfrei} (smoke-free) which should decrease the rate of smokers, see also \cite{bmg-rauchfrei}. Amongst others, an evaluation and good summary of the intermediate results can be found, e.g., in Wenig et al. \cite{wenig}.
\par
In this article we are interested in the effectiveness of smoking cessation with the help of hypnosis, a classical  method for quitting smoking. Typically, smoking intervention by hypnosis uses a concept where people meet a hypnotherapist in person for one or more sessions. During the COVID-19 pandemic meetings in person were restricted in a very extended way laws. In many areas of daily life meetings went \textit{online}. While online hypnosis was offered by a rather limited number of hypnotists before the pandemic, it seems that nowadays many hypnotists started to offer online hypnosis sessions  for smoking cessation or weight loss. During an online session via online conference tools like Skype, Zoom, etc., the patients stay at home and turn on their webcams, and the hypnotist may perform his session almost in the same way as in a real-life session. Nonetheless, there are obviously still differences between online hypnosis sessions and sessions in real life. On the one hand side, the direct contact between hypnotist and subject is lost, which may reduce the chance of success; on the other hand side, subjects are staying in their comfort zone at home, where they may relax even easier than in an external room provided by the hypnotist. This leads to the question of effectiveness of online hypnosis, and
this question was the starting point for our own field study, in which we want to evaluate the \textit{effectiveness of online hypnosis for smoking cessation}.
For this purpose, we made a call for smokers wishing to stop smoking by online hypnosis intervention. Our hypnosis intervention protocol consists of two online hypnosis sessions and individual post-session treatment by audio files. Effectiveness is examined by questionnaires after each session, and additionally after three and six months. The protocol will be described in detail in Section \ref{sec:method}.
At this point we remark that \textit{we neither  aim on comparing the success rate of hypnotic intervention with other intervention types nor we want to equate this report with a fully pre-designed study in a narrower sense}. The aim of this article is to summarise our observations and to statistically describe the data which we have collected. In particular, we have the following goals according to our observations:
\begin{itemize}
\item estimate the \textit{success rate by intervention via online hypnosis}, 
\item determine the \textit{main criteria which influence the success rate}, 
\item and detect \textit{possible problems} which may lead more likely to flops. 
\end{itemize}
Let us provide some results on effectiveness of smoking intervention by hypnosis. Barrios \cite{barrios} summarized the results of several studies concerning effectiveness of different intervention methods for smoking cessation. Averaging over several studies, where participants received hypnosis intervention for smoking cessation (six sessions in average), he came to a success rate of $93\%$. Other therapies seem to be much less effective: $72\%$ of smokers, who received a behavioural therapy, were successful after 22 sessions in average, while only $38\%$ of smokers, who received a psychoanalytic treatment over several years, were successful after 600 sessions in average. Amongst others, we want to mention the following articles which evaluated the success rate of hypnotic intervention for smoking cessation:
Johnson and Karkut \cite{johnson-karkut} came to a $86.5\%$ success rate after three months when combining hypnosis and aversion techniques. In particular, there was no gender-specific difference in the success rate.  Barber \cite{barber} calculated a success rate of $90.6\%$ when using an integrated approach combining hypnotic methods and rapid smoking treatment protocol. The article of Elkins and Rajab \cite{elkins2004} presents preliminary data of a three-session hypnosis intervention for smoking cessation, which comes close to the method of hypnotic intervention used in our article; they came to a success rate of $81\%$. However, the picture of effectiveness of hypnotic intervention for smoking cessation is still mixed: e.g.,
the article of Ahijevych, Yerardi and Nedilsky \cite{ahijevych}  ended up with a success rate of less than $25\%$.  Nonetheless, the critical article of Green and Lynn \cite{green-lynn} compared $59$ studies and they came to the conclusion to classify hypnosis as a ``possibly efficacious'' treatment for smoking cessation. Concerning smokers who try to quit on their own (so-called \textit{self-quitters}) we want to mention the following articles: Hughes et al. \cite{hughes04} summarised that $3\%-5\%$ of self-quitters are successul after $6-12$ months, and  Hughes et al. \cite{hughes92} reported  that only $33\%$ of motivated self-quitters remain smoke-free after two days of starting to quit.
\par
This paper is organized as follows: in Section \ref{sec:subjects} we explain how we found participants and we give a short statistical overview on the composition of the participants. In Section \ref{sec:method} we present the protocol of our study, including a detailed description of the methods used for the hypnotic intervention for smoking cessation. Finally, in Section \ref{sec:results} the main observation and evaluation results of our study are presented, and the results and problems observed are discussed in Section \ref{sec:discussion}.

\section{Subjects}
\label{sec:subjects}

The participants for this field study were acquired by calls in social media (Facebook, Twitter, Instagram), regional newspapers (Dorstener Zeitung, Passauer Neue Presse, Grazer Woche), and via radio channels (UnserRadio, Radio Galaxy). Finally, $83$ subjects from Germany, Austria and Switzerland were taken into account for this study, consisting of $49$ ($59\%$)  female and $34$ ($41\%$) male subjects. All interested subjects were accepted for participation unless severe psychological problems in the past were given, which led to an exclusion of participation. 
The participants had to pay a small nominal fee in order to cover the study's expenses (for this study, \textit{no} fundings were  requested)  and to prevent participation of people, who just want to ``try'' without any serious intention. 
\par
The average age of the participants was $43$ (median: $41$), while the minimum age was $23$ and the maximum age was $70$ with a standard deviation of $11.2$. Furthermore, $48\%$ of the participants were married, while $18\%$ were divorced. We also remark that $36\%$ of the participants were suffering from overweight.
\par
\begin{figure}[h]
\includegraphics[scale=.25]{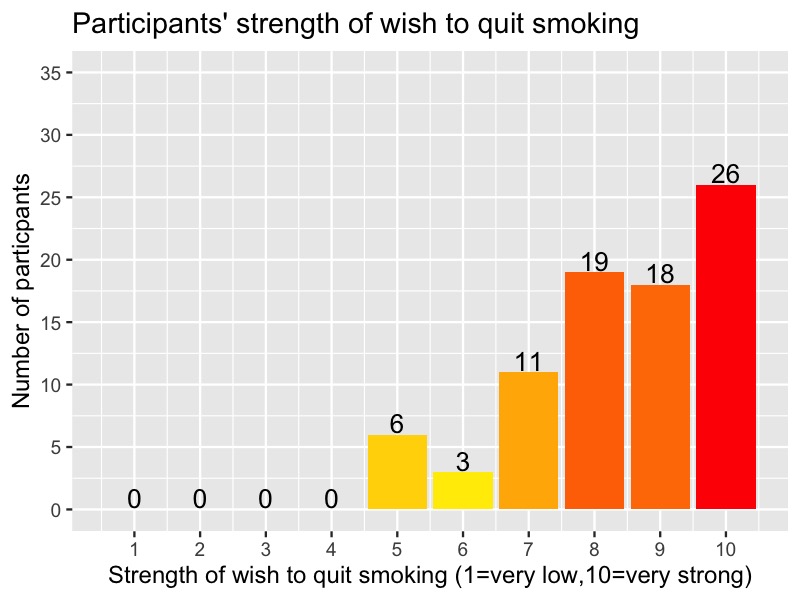}
\hspace{0.5cm}
\includegraphics[scale=.25]{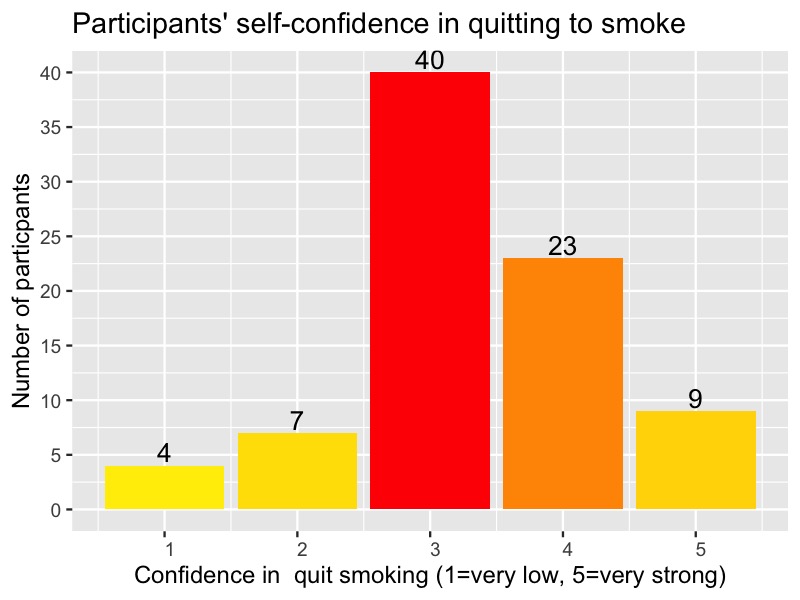}
\caption{Participants' strength of wish and self-confidence to quit smoking.}
\label{fig:wish}
\end{figure}

The average daily amount of cigarettes smoked by the participants \textit{before} the intervention was $18.2$ cigarettes per day (median: $18$), while the minimum was $2$ cigarettes per day and the maximum was $50$ cigarettes per day. The average number of years of smoking was $24.4$ years (median: $22$), with a range between $5$  and $50$ years. The participants' strength of wish to stop smoking and their self-confidence in the intervention is sketched in Figure \ref{fig:wish}. 
As one can see, the majority has a strong or very strong wish to quit smoking; nonetheless, around $18$ subjects (strength of wish at level $5$ or $6$ or self-confidence level $1$ or $2$) seem to participate without having a strong desire to stop smoking and/or have rather less self-confidence in the intervention. One third of the particpants have an intermediate confidence in the hypnosis intervention. It is clear that the subject's collaboration in some way is indispensable for the success, but we want to study the main connections between sufficient collaboration/motivation and success.
\par
We also remark that  $92\%$ among the $83$ participants reported that they have failed in previous unassisted attempts to quit smoking permanently. Moreover, $24\%$ of the participants already had general experience with hypnosis. 
\par
Finally, we want to give insights in which situations the subjects were smoking before the intervention. Each participant had to give a grade (on a scale from $1$ to $5$) for a given situation, in which he smokes very often (grade $1$) or very rarely (grade $5$).
Figure \ref{fig:smoking-situations} lists the favourite reasons, where participants gave grade $1$:
\begin{figure}[H]
\includegraphics[scale=.3]{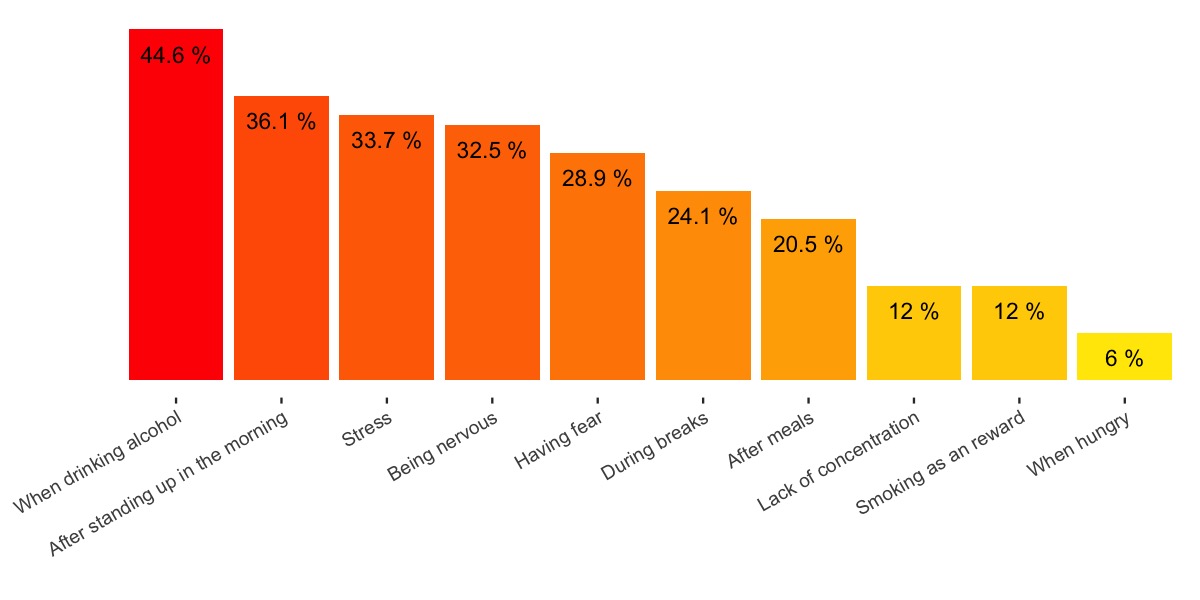}
\caption{Participants' main situations for smoking.}
\label{fig:smoking-situations}
\end{figure}
E.g., for $44.6\%$ of the participants it is quite hard \textit{not} to smoke when consuming alcohol. Furthermore,  $58\%$ of the participants tend to smoke when drinking alcohol, after standing up in the morning or when having stress. 
%

\begin{figure}[H]
\includegraphics[scale=.24]{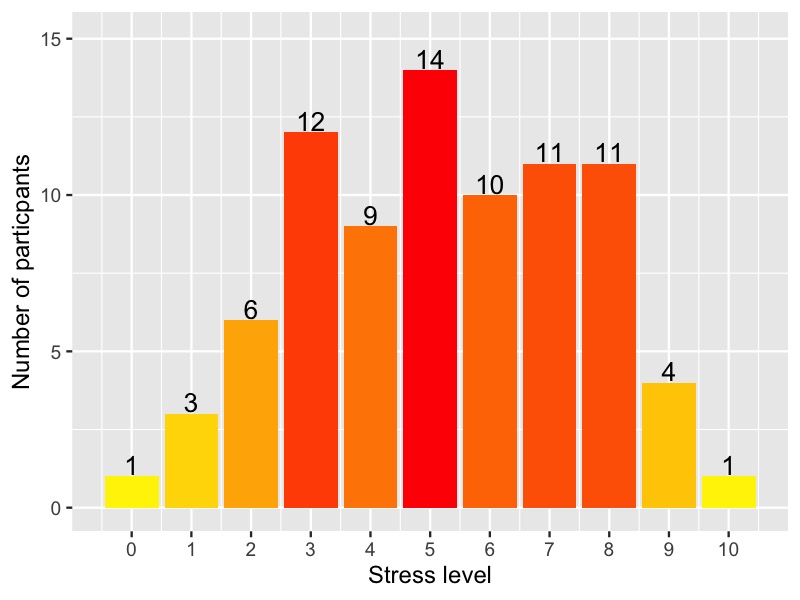}
\hspace{0.1cm}
\includegraphics[scale=.24]{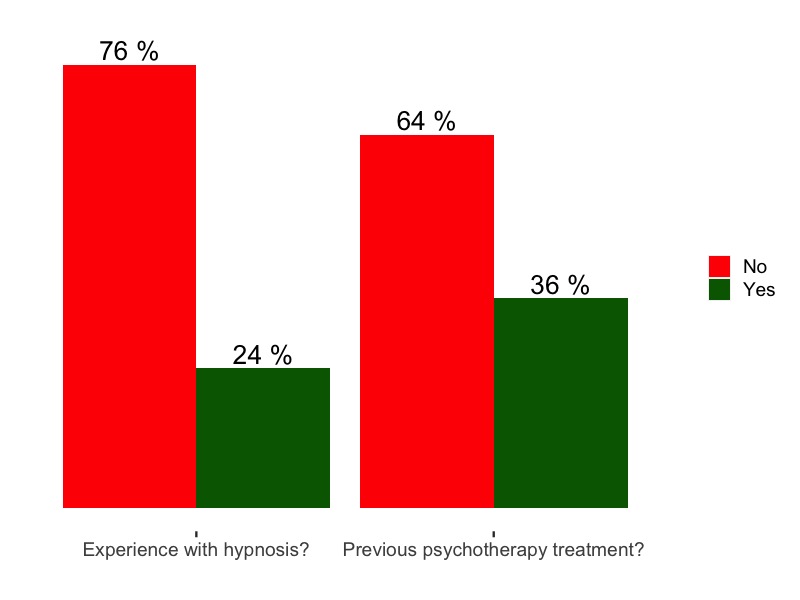}
\caption{Participants' stress level (0=very low, 10=very strong) and experiences with hypnosis or psychotherapies.}
\label{fig:stresslevel}
\end{figure}

The participants were splitted up into different groups according to their stress level, psychological problems in the past (these persons received an induction leading into a lighter state of trance in order to avoid any problems occuring from the past) and other secondary factors like the ones mentioned above. The distribution of the participants' stress level and previous experience with hypnosis and/or psychotherapies is sketched in \mbox{Figure \ref{fig:stresslevel}.}

Although we tried to de-randomize the group structure, we mention that it was only possible in a very limited way to adapt the suggestions given to the different groups during  the hypnosis sessions.
\par
Finally, we remark that we even had $15$ participants more, who attended the first hypnosis sessions but did \textit{not} send in the questionnaire after this session or who did \textit{not} attend the second session; we also had $16$ participants who attended -- from an objective point of view --  the hypnosis session in an inappropriate manner (e.g, the started to pet their dogs/cats during the session). Those particpants were excluded from the evaluation.

\section{Method}
\label{sec:method}

In this section we describe the intervention protocol, which consists of the following four main points: introduction to hypnosis, a first hypnosis session, individual treatment with MP3 audio files and a second hypnosis session. We will explain these stages in detail in the following and we also give a comment on the data collection. Due to the COVID-19 pandemic, all hypnosis sessions took place online via \textit{Zoom}. The participants were splitted up into different rather homogeneous groups (minimal group size was $5$, maximum group size was $20$) by taking into account different stress level, psychological problems in the past and other secondary factors.
\par
All online meetings and hypnosis sessions were held by the authors of this article, who are both certified hypnocoaches.

\subsection{Introduction to Hypnosis}

All participants were invited to join an online meeting via Zoom, where we explained our study, formulated the goals and gave an explanatory introduction to hypnosis. It was our main goal to reduce or even eliminate false opinions on hypnosis (e.g., originating from TV shows etc.) and to explain what happens during a hypnosis session. This education was very important in order to prepare all participants as good as possible. Moreover, all participants received guidelines for the rules of conduct for the hours before and after the hypnosis sessions (e.g., we remarked that people should not drink coffee just before the sessions or have to strictly avoid testing after the session if smoking still ``works'').

\subsection{Session 1}

In the first session we used a classical hypnosis approach for anti-smoking intervention consisting of the following parts:
\begin{enumerate}
\item Classical induction with encouraging people to quit smoking.
\item $7+2$ deepening: this deepening uses the approach that people, in general, can only think of $7$ plus or minus $2$ things at the same time. Subjects are requested to think at different parts of the body step by step. This should overload the consciousness and help to improve the relaxation.
\item Dissociation of mental blockades: this part aims on preparing the subjects to get rid of mental problems  and to become open for the following part of suggestions. The subjects went on a virtual  journey to a waterfall, under which all problems and troubles in their mind and soul could be washed away.
\item Suggestion booster: this part shall boost the upcoming suggestions.
\item Commitment for stopping to smoke: in this part the subjects are prepared to stop smoking; they are encouraged to get rid of smoking.
\item Posthypnotic suggestions: in this part the main suggestions are given which aim on the non-necessity of smoking, better feelings and emotions without cigarettes, and so on. Suggestions were kept quite general in order to reach all participants.
\item Integration of the goal: after having implemented the anti-smoking suggestions some further suggestions were given in order to deepen the suggestions and to intensify them.
\item Closing the session: people were led out of trance.
\end{enumerate}
The presented structure of this session follows a well-known scheme which is widely used by hypnotherapists for smoking cessation.

\subsection{MP3 Audio Files \& Post-Session Treatment}

Three days after the first session all participants received two MP3 files for individual treatment and intensification. The first audio file is a classical $15$-minutes short hypnosis with suggestions for deepening the hypnotic anti-smoking suggestions. Furthermore, the participants got a subliminal MP3 file with anti-smoking suggestions. The participants were encouraged to listen to these MP3's whenever  the impulse to smoke increases and as often as they wanted (e.g., during lunch breaks, in the evening, etc.), but \textit{at least once per day during the first two weeks} unless the impulse to smoke is already gone. This task is an important part of the intervention protocol, which is requested to be followed by the participants.

\subsection{Session 2}

The second hypnosis session took place \textit{one week after the first session}. It consisted of the following parts:
\begin{enumerate}
\item Classical induction with eye fixation.
\item Brain twister deepening: this deepening sequence aims on creating some confusion, which overloads the subject's consciousness resulting in an even deeper state of relaxation.
\item Dissociation of mental blockades:  analogously to the first session the subjects were asked to  imagine an ice block consisting of all problems and troubles in their mind, which is then melted away.
\item Suggestion booster: this part shall boost the upcoming suggestions.
\item Commitment for stopping to smoke and reviewing the personal results so far: subjects were encouraged to look back on their (maybe partial) success and to improve their trust in a successful smoking cessation.
\item Posthypnotic suggestions: these suggestion aim once again on better feelings and emotions without cigarettes, and so on. 
\item Integration of the goal: some further suggestions were given in order to deepen the anti-smoking suggestions and to intensify them.
\item Closing the session: people were led out of trance.
\end{enumerate}
The structure of this session follows also a well-known scheme which is widely used by hypnotherapists for a second hypnosis session in the context of smoking cessation. We note that the second session should \textit{not} be seen as a ``second chance'', but as an intensification of the first session which should lead to a sustainable smoke-free life. The participants were informed about this fact during the introduction event.

\subsection{Data Collection}

The participants were asked to submit questionnaires at the follwing stages:
\begin{enumerate}
\item Before the first session for getting information about the smoking behaviour and to exclude psychological problems.
\item Three days after Session $1$.
\item One week after Session $2$, that is, two weeks after Session $1$.
\item Approx. six weeks after Session $2$.
\item Approx. six months after Session $2$.
\end{enumerate}

\section{Results}
\label{sec:results}

In this section we present the main observations and evaluation results of the collected data from our study which is structured as follows. First, we split up the participants into two groups identified according to the observations we have made during the intervention process (in particular, during the hypnosis sessions). Afterwards we present the success rates of the anti-smoking intervention considered from different points of view. Finally, we compare the groups and give insights into each group, where we detect some factors which are in correlation with success or failure of the intervention.

\subsection{Individual Participation Related Group Assignment}
\label{subsec:groups}

One of the main observations during the intervention process was that the subjects attended the hypnosis sessions quite differently. While there was a group of subjects who really wanted to quit smoking and put effort into the anti-smoking process, there was another group of subjects who did \textit{not} seem to participate with sufficient effort by following our intervention protocol. It turned out that a good indicator for the degree of following our guidelines is given by the \textit{intensity of the individual post-treatment} starting after Session $1$. The participants were asked to listen to the audio files \textit{at least once per day} for the first two weeks after Session $1$ (unless there is no impulse to smoke any more), depending on how difficult it is to resist smoking. While there were many subjects, who listened to the audio files almost daily, also many subjects did listen to the audio files only very rarely. We have identified this difference as a good indicator whether subjects participated in a reasonable motivated manner, and therefore we classify the participants accordingly: since there were $12$ days between the receipt of the audio files and day $14$ after Session $1$, we classify those participants as \textit{motivated} who finally quitted smoking after Session $2$ and are still non-smoker after six months or who have listened to the audio files at least $12$ times during the first two weeks. In other words, those participants were immediately successful after Session $2$ and/or they have at least tried to put enough own effort into the intervention process.
According to our observations we may classify the subjects \textit{a-posteriori} into one of the following two disjoint groups:
\begin{itemize}
\item \textit{Lazy group:}  this group contains participants, who did attend the hypnosis session in a mostly serious manner but did \textit{not} follow the intervention protocol as requested.
\item \textit{Motivated group:}  participants, who did attend the hypnosis session in a serious manner and did follow the intervention protocol as requested.
\end{itemize}
The group sizes are as follows:
\begin{center}
\tablinesep=2ex\tabcolsep=10pt
\begin{tabular}{|c|c|c|}
\hline
Group & Number of participants & Percentage\\
\hline
\hline
\textit{Motivated group} & $28$ & $33.7\%$ \\
\textit{Lazy group} & $55$ & $66.3\%$\\
\hline
Total & $83$ & $100\%$\\
\hline 
\end{tabular}
\end{center}

For the evaluation of the effectiveness of online hypnosis for smoking cessation the \textit{Motivated group} is essentially most important. 
In Section \ref{subsec:motivated-vs-lazy} we demonstrate the impact when separating   the \textit{Motivated group} and the \textit{Lazy group} by a lower number of minimal audio files listenings, and in Sections \ref{subsec:discussion-decomposition} and \ref{subsec:discussion-M-L} we discuss the question of sufficient motivation regarding the separation between the \textit{Motivated group} and the \textit{Lazy group}.

\subsection{Success Rates}
\label{subsec:success-rates}

\begin{figure}[H]
\includegraphics[scale=.3]{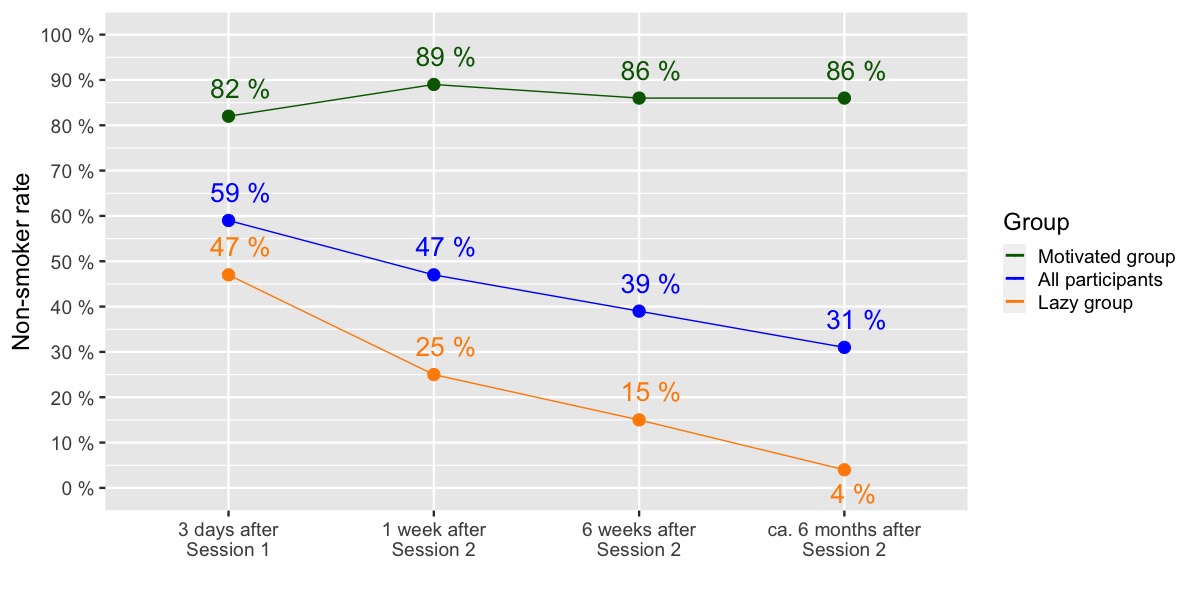}
\caption{Evolution of non-smoker rates by group.}
\label{Nonsmoker-lazy-main}
\end{figure}

All participants reported three days after Session $1$ whether they have smoked since Session $1$. One week after Session $2$ they reported whether they have smoked since Session $2$. Approx. six weeks and approx. three months after Session $2$ the participants reported whether they are still/again smoking or not. The evolution of non-smoker rates for both groups are sketched in Figure \ref{Nonsmoker-lazy-main}. 

We remark that two participants in the \textit{Lazy group} are \textit{not} smoking after six months.

\subsection{Amount of Cigarettes Consumption}
\label{subsec:amount-cigarettes}

The main goal of the anti-smoking hypnosis intervention was, of course, to eliminate tobacco consumption. Nonetheless, we take a closer look on the smoking behaviour after six months in the both groups \textit{among smokers}. The average amount of daily cigarettes before the intervention process and after six months \textit{among smokers}  is summarized by group  in the following table, where the corresponding standard deviation is noted in brackets:

\begin{center}
\tablinesep=2ex\tabcolsep=10pt
\begin{tabular}{|c|c|c|}
\hline 
Group & Mean before intervention  & Mean after $6$ months  \\
\hline
\hline
\textit{Motivated group} & $21.8$ ($9.6$) & $14.8$ ($17.3$)\\
\textit{Lazy group} & $18.7$  ($6.2$) &  $12.6$ ($6.2$)\\
\hline
All groups & $18.9$ ($6.4$) & $12.8$ ($7.2$)\\
\hline
\end{tabular}
\end{center}

\begin{figure}[h]
\includegraphics[scale=.3]{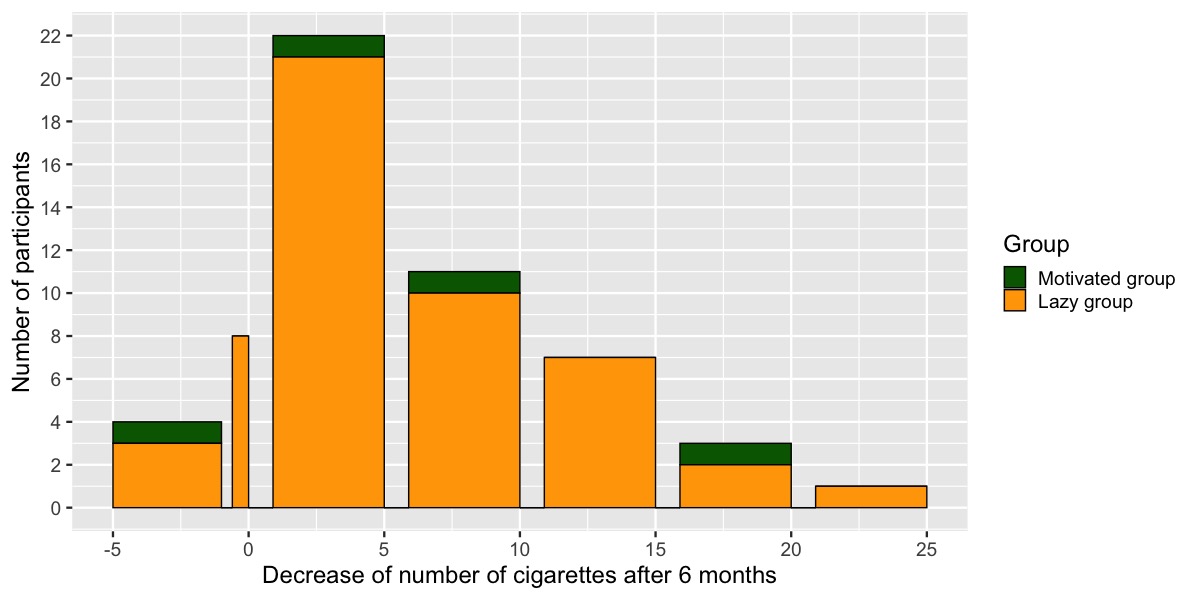}
\caption{Decrease in the average number of daily cigarettes among smokers after six months.}
\label{cigarettes-decrease}
\end{figure}

The differences of the individual  average number of daily cigarettes \textit{before the intervention} with the individual average number of daily cigarettes \textit{after six months} are sketched in Figure \ref{cigarettes-decrease}.

Among all smokers we observe an average decrease of $6.2$ cigarettes per day after six months.
\par

\begin{figure}[H]
\includegraphics[scale=.35]{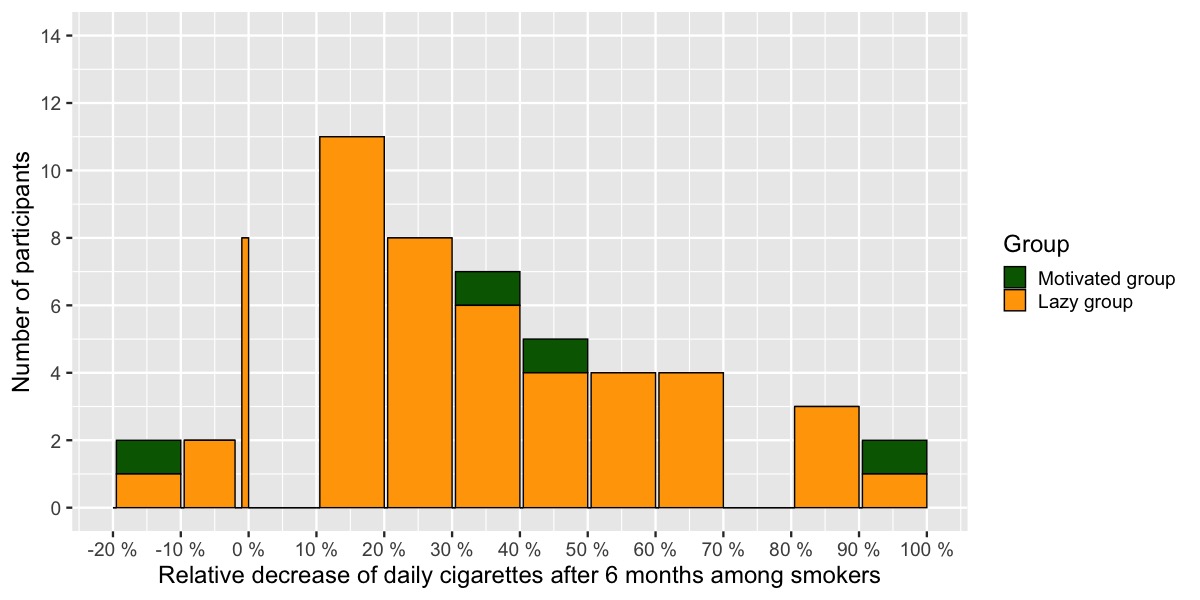}
\caption{Relative decrease in daily cigarettes among smokers after six months.}
\label{cigarettes-decrease-prozentual}
\end{figure}

The \textit{relative decrease} in daily cigarettes consumed before the intervention and after six months \textit{among smokers} is sketched in Figure \ref{cigarettes-decrease-prozentual}.

\begin{center}
\begin{figure}[H]
\includegraphics[scale=.3]{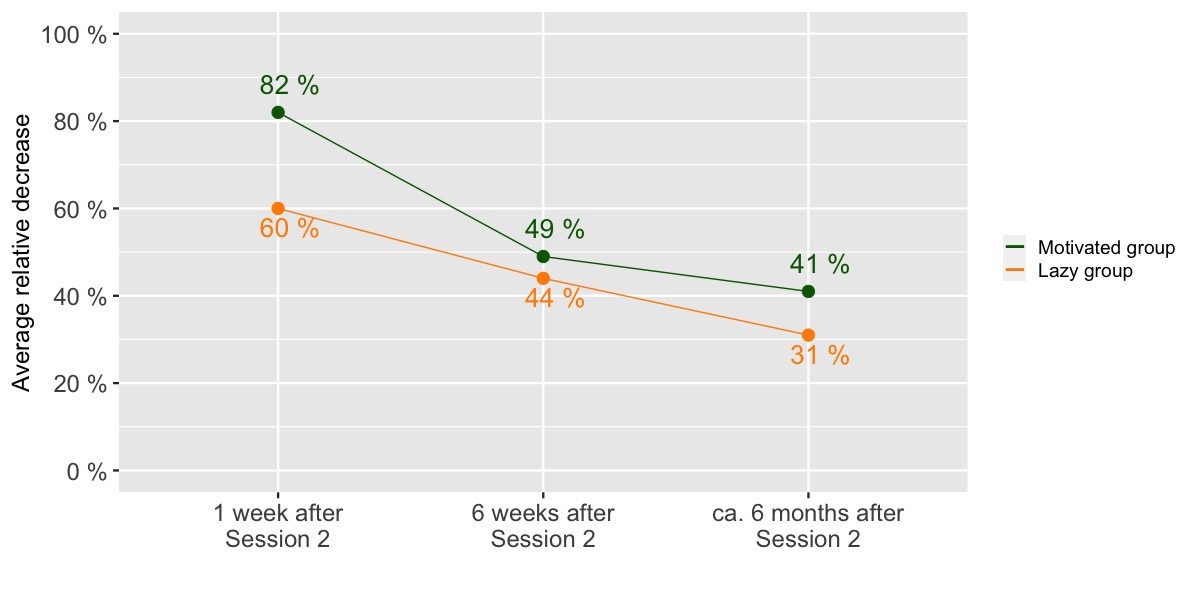}
\caption{Average decrease of daily cigarettes  \textit{among} smokers.}
\label{cigarettes-average-evolution}
\end{figure}
\end{center}

The evolution of the average relative decreases in daily cigarettes \textit{among smokers} is sketched in Figure \ref{cigarettes-average-evolution}.

The  boxplots in Figure \ref{cigarettes-decrease-prozentual-boxplot} compare the location of the individual relative decreases by group:

\begin{figure}[h]
\includegraphics[scale=.25]{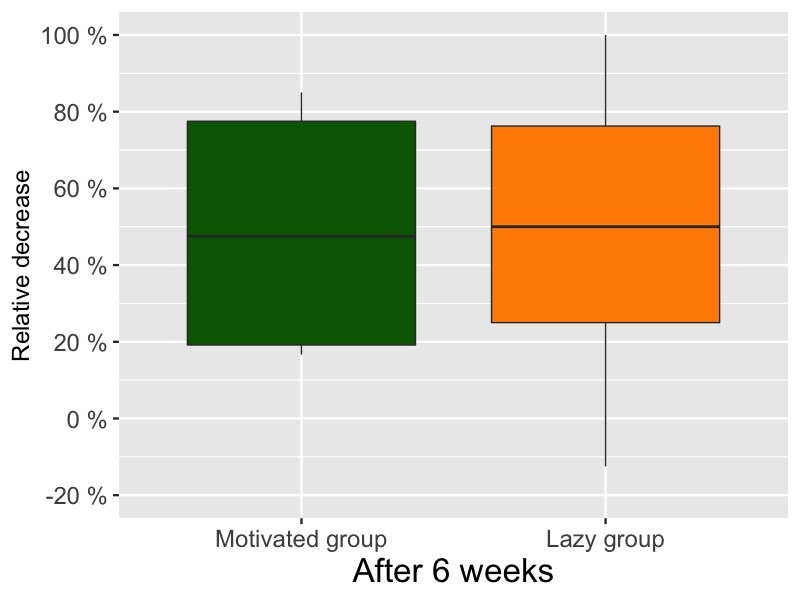}
\hspace{0.5cm}
\includegraphics[scale=.25]{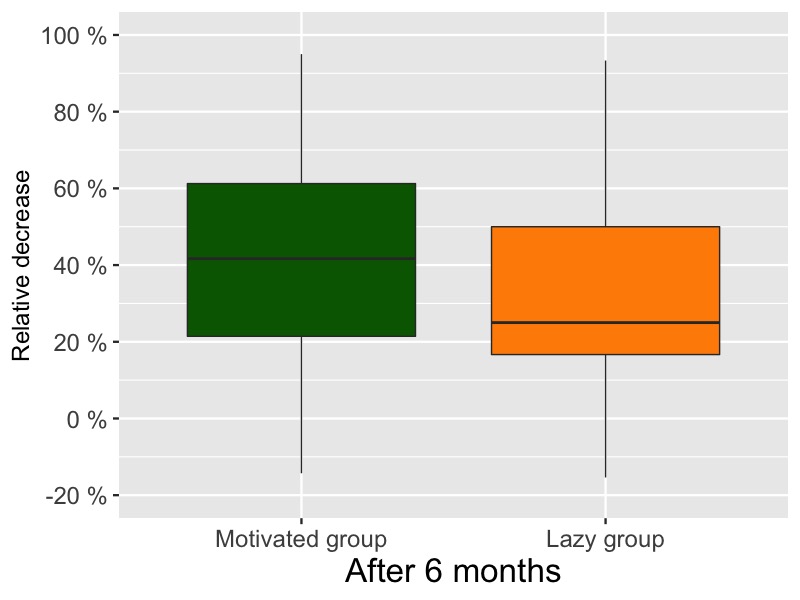}
\caption{Relative decrease in daily cigarettes among smokers.}
\label{cigarettes-decrease-prozentual-boxplot}
\end{figure}

Among all participants who are still smoking after six months we have an average decrease of $32 \%$ in tobacco consumption.

\subsection{Group Comparision}

\subsubsection{Self-confidence and Strength of Wish}
\label{subsec:self-confidence}

We compare two important \textit{a-priori} indicators for the motivation of the participants: the self-confidence score and the strength of wish to become non-smoker; see Figure \ref{fig:wish}.
\par
Comparing the self-confidence score by group on a scale from $1$ (very low) to $5$ (very high) 
\textit{before} the first hypnosis session  (see Figure \ref{fig:wish}) yields the following picture:

\begin{center}
\tablinesep=2ex\tabcolsep=10pt
\begin{tabular}{|c|c|c|}
\hline 
Group & Mean & Standard deviation \\
\hline 
\hline
\textit{Motivated group} & $3.46$ & $0.96$ \\
\textit{Lazy group} & $3.24$ & $0.94$ \\
\hline
\end{tabular}
\end{center}

The strength of wish score on  a scale from $1$ (very low strength) to $10$ (very high strength) from Figure \ref{fig:wish} by group is summarised in the following table: 

\begin{center}
\tablinesep=2ex\tabcolsep=10pt
\begin{tabular}{|c|c|c|}
\hline 
Group & Mean & Standard deviation \\
\hline
\hline
\textit{Motivated group} & $8.18$ & $1.42$ \\
\textit{Lazy group} & $8.55$ & $1.54$ \\
\hline
\end{tabular}
\end{center}

The self-confidence \textit{after} Session $1$ -- now on a scale from $1$ (very low) to $10$ (very high) -- is as follows:

\begin{center}
\tablinesep=2ex\tabcolsep=10pt
\begin{tabular}{|c|c|c|}
\hline 
Group & Mean & Standard deviation \\
\hline
\hline
\textit{Motivated group} & $7.75$ & $2.10$ \\
\textit{Lazy group} & $6.36$ & $2.41$ \\
\hline
\end{tabular}
\end{center}

We remark that the scale change in the self-confidence score was made in order to get better insights, although comparing the scores before and after Session $1$ is now a bit vague.

\subsubsection{Motivated Group vs. Lazy Group}
\label{subsec:motivated-vs-lazy}

\begin{figure}[H]
\includegraphics[scale=.28]{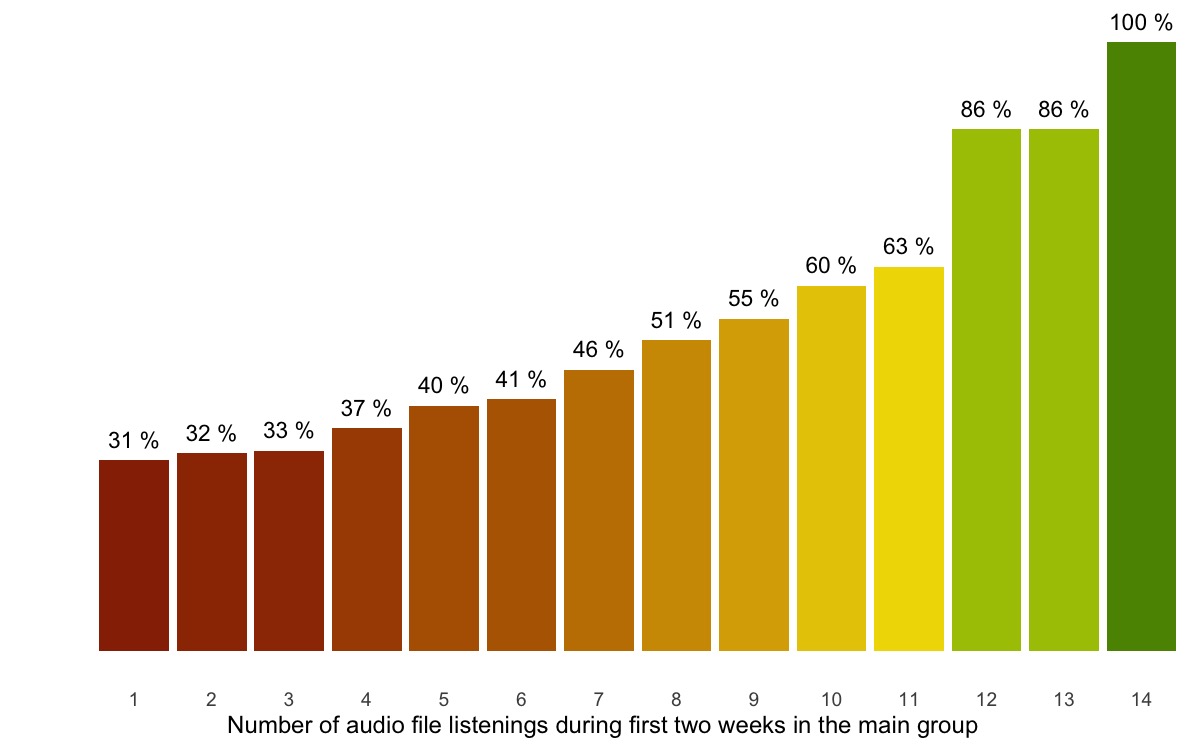}
\caption{Dependency of the success rate on the number of audio file listenings during the first two weeks.}
\label{fig:success-listenings}
\end{figure}

We separated the subjects in the \textit{Motivated Group} and the \textit{Lazy Group}  by considering the number of audio listenings within the first two weeks of the intervention. 
The participants were requested to listen to the audio files at least once per day  during this period, hence at least $12$ times. A legitimate question is now what happens if we replace this critical number by a lower number. If we put a participant into the \textit{Motivated group} if he/she has \textit{not} smoked since Session $2$ any more (maybe with a lower number of listenings because it was not hard for them to resist smoking) or have listened to the audio files at least $N$ times, then we obtain a non-smoker rate after six months depending on $N$ as sketched in \mbox{Figure \ref{fig:success-listenings}.}
Observe that the hightest jump occurs at $N=12$.

\subsubsection{Cigarette Taste}
\label{subsub:compare}

\begin{center}
\begin{figure}[h]
\includegraphics[scale=.30]{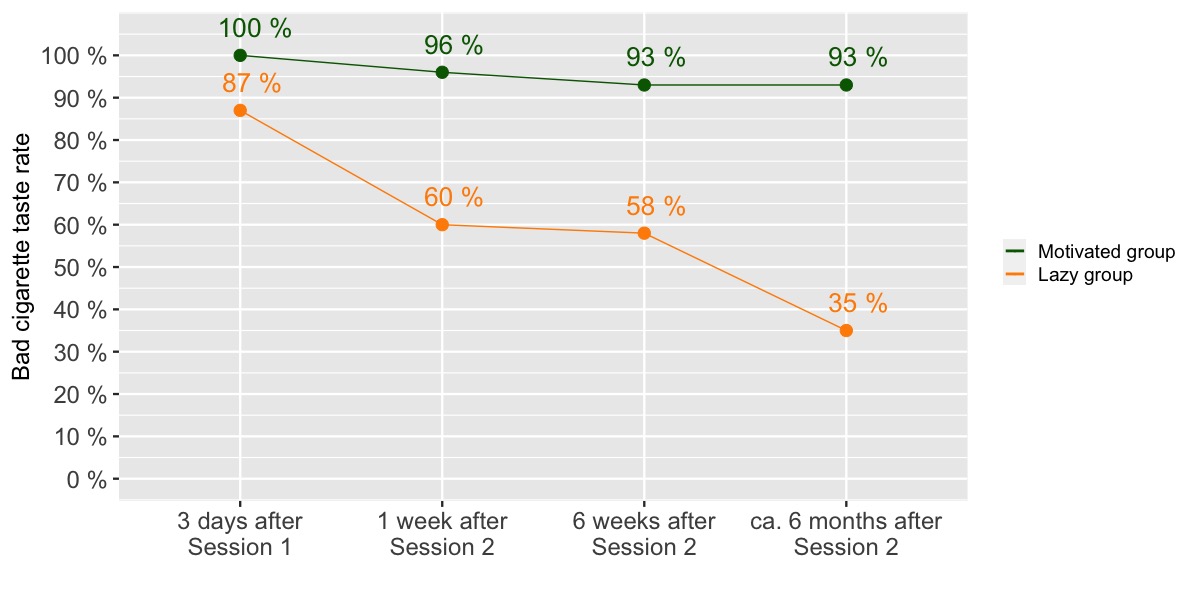}
\caption{Rate of participants who do \textit{not} smoke or for whom cigarettes are tasting bad.}
\label{cigarettes-bad-taste}
\end{figure}
\end{center}

The participants were asked at different stages of the intervention whether cigarettes taste the same as before the intervention or not. The results are shown in  \mbox{Figure \ref{cigarettes-bad-taste}.}

\subsection{Group Specific Observations}

In this section we investigate group specific criteria for success or failure of the smoking intervention. We use Fisher's exact test in order to detect possible stochastic dependencies of different factors with the success.

\subsubsection{Motivated Group}

Consider the main situations for smoking listed in Figure \ref{fig:smoking-situations}. Among the main reasons only the combination of smoking when hungry seems to have a negative effect in this group: independence of success after six months and smoking habits when hungry can be rejected with a $p$-value of $0.011$. Furthermore, we detect some light
 negative dependency with being nervous or having fear before the hypnosis sessions with a $p$-value of $0.094$.
Since there are only four smokers after six months in this group, the statistical dependencies are a bit uncertain and should be taken with caution only.

\subsubsection{Lazy group}
\label{subsec:lazy}

In this group we consider the success of being abstinent \textit{after the first two weeks}. We can \textit{reject} stochastic independence of the success after two weeks with the following factors:
\begin{center}
\begin{tabular}{|c|c|}
\hline
Factor & $p$-value \\
\hline
\hline
Smoking when hungry & $0.012$\\
Smoking when having lack of concentration & $0.014$\\
Individual stress level & $0.015$ \\
After standing up in the morning & $0.040$ \\
Relationship & $0.037$\\
Influence from other persons &  $0.054$ \\
Smoking when being angry & $0.074$ \\
\hline
\end{tabular}
\end{center}

\subsection{General Summary}
\label{subsub:general-result}

From Subsections \ref{subsec:success-rates} and \ref{subsec:amount-cigarettes} we can draw the following summary: we observe  that $85\%$ among \textit{all} participants do \textit{not} smoke after six months or have at least reduced their daily tobacco consumption.

\section{Discussion}
\label{sec:discussion}

\subsection{Summary of the Study}
This study had the aim to investigate the effectiveness of smoking cessation via \textit{online hypnosis} in small groups, which has \textit{not} been studied as far as the authors know. For this purpose, participants were acquired by an international call over social media, regional newspapers and radio channels. All interested applicants were allowed to participate as long as there were no contraindications (e.g., due to psychological problems in the past).
The participants received two online hypnosis sessions in small groups and also audio files for individual post-treatment and intensification. Furthermore, they were briefed on the intervention process and their required collaboration in advance.
 At different stages the participants reported whether they are still abstinent of cigarettes and tobacco or not. 
\par
According to the subjects' willingness/motivation to contribute to the smoking cessation process we have identified two groups of participants:
a group with rather less or intermediate level of motivation which did \textit{not} satisfy the expected level of collaboration (called  \textit{Lazy group}), and a group where the motivation satisfied our intervention protocol (called  \textit{Motivated group}). The study resulted in a $86\%$ abstinency rate in the  \textit{Motivated group} after six months, while most of the participants in the other group are  smoking again after $6$ months. Moreover, \textit{among all smokers} there is an average decrease of $32\%$ of tobacco consumption after six months since the beginning of the intervention. 

\subsection{Decomposition of Participants}
\label{subsec:discussion-decomposition}

Let us discuss the decomposition of the participants into different groups regarding their motivation. At first glance,  the  subjects  formed  a quite unclear picture. Our observation was that some subjects participated well-motivated while others didn't. This was the starting point to consider motivated subjects separately and to detect an objective indicator which separates subjects according to their motivation.
The assignment of these subjects either to the \textit{Motivated group} or to the \textit{Lazy group} was, of course, less obvious at the beginning. Therefore, we demanded for members to be in the \textit{Motivated group} that our guidelines (in particular, to listen to the audio files at least once per day during the first two weeks after Session $1$) are followed  during the whole intervention process. 
\par
For this purpose, we have found an objective indicator by considering the number of audio listenings within the first two weeks of the intervention. Continuous post-session treatment by listening to audio files is a good support to strengten the abstinence of tobacco and to deepen the suggestions received during the hypnosis sessions. From this point of view, participants, who listened to the audio files rarely during the first two weeks of intervention, obviously seemed  to have some lack of motivation. 
Setting the critical number of audio listenings to $12$ is justified by different reasons: first, $12$ listenings correspond to an average number of listenings of once per day during the first two weeks; second, Figure \ref{fig:success-listenings}  illustrates how the number of listenings influences the success rate, where the highest jump occurs at $N=12$, a reasonable change point to choose. 
\par
From our point of view, subjects in the \textit{Lazy group}  seem to be rather to lazy to follow the complete intervention protocol; instead, they maybe hoped that the hypnosis sessions does all the work for them and no own effort is required in order to strengthen the effects of hypnosis and to complete the intervention process, while most of the people in the \textit{Motivated group} seemed to have a stronger motivation. 
We note that not all non-smokers after six months in the \textit{Motivated group} listened to the audio files at least $12$ times; these participants were already completely abstinent of tobacco immediately after Session $2$, since the hypnosis seemed to work very well for them. Of course, motivation cannot be classified by the number of audio listenings solely, but -- in absence of other reasonable, objective indicators -- this number is at least one obvious indicator for the degree of  the subjects' personal motivation and collaboration. One critical point, of course, is that we have an \textit{a-posteriori} group decomposition and there is the question which factors influence the effectiveness and motivation beyond the individual audio listenings. Further work has to be done in this direction. For a discussion regarding differences between the \textit{Motivated group} and the \textit{Lazy group}, we refer to Subsection \ref{subsec:discussion-M-L}.
\par
Recall that we have split up the participants into small groups for the single hypnosis sessions. We note that we did \textit{not} observe any different behaviours concerning which hypnosis session has been visited by the participants. 
\par
Let us also remark that our big call for participants together with the very low  participation fee of $25$\,\euro{}  (we note that hypnosis intervention of our proposed form starts from approx. $200$\,\euro{} in Germany) presumably led to participation of many people, who rather wanted to get the experience of hypnosis without putting too much \textit{own} effort into the whole intervention process. This could be an explanation why the \textit{Lazy group} has $51$ members and the \textit{Motivated group} consists of $28$ members only.
\par
Overall, the presented reasoning underlines the necessity  of motivation and collaboration which led to the presented decomposition of the participants into  different groups. 

\subsection{Success Rates}

As one can see in Figure \ref{Nonsmoker-lazy-main} the motivation and willingness to put some own effort into the intervention process is \textit{essential} for the success to become non-smoker. In particular, observe the higher rate of non-smokers in the \textit{Motivated group} one week after Session $2$. We interpret this higher rate after Session $2$ as the necessity of an additional hypnosis session for deepening the suggestions and improving the effects: some smokers were not able to stop smoking completely after Session $1$ but Session $2$ let them finally stop.
 However, the non-smoker rates in the \textit{Lazy group} are decreasing quickly and are already quite low after six weeks. 
 \par
Our main conclusion is that the hypnosis sessions in small groups lay the base for a successful smoking intervention but  some individual post-session treatment is inevitable in most cases for a sustainable effect. In particular, sufficient motivation and putting some own effort into the intervention process is indispensable in most cases. 

\subsection{Amount of Daily Cigarettes}
 
An astonishing evaluation result is the significant decrease in the average number of daily cigarettes as reported in Subsection \ref{subsec:amount-cigarettes}. Although the hypnosis intervention did in most cases \textit{not} lead to tobacco abstinence for the participants in the \textit{Lazy group}, we observe at least some ``partial success'' in a noteworthy decrease in the amount of daily cigarettes consumed. In average, the participants who still smoke after six months have reduced their tobacco consumption by $32\%$.
We remark that it is rather unclear whether the  decrease in daily cigarettes in the \textit{Lazy group} arises as some partial hypnotic effect or as ``solidarity''.  
We compare and discuss the groupwise decrease in daily cigarettes in the upcoming sections in more detail.

%

\subsection{\textit{Motivated Group} vs. \textit{Lazy Group}}
\label{subsec:discussion-M-L}

Let us discuss the different evolution in both groups. First of all, the success rates in Figure \ref{Nonsmoker-lazy-main} demonstrate a significant different behaviour: while the number of non-smokers in the \textit{Motivated group}  is very high and increasing or constant on a high level, the number of non-smokers in the \textit{Lazy group} is monotonously decreasing  already after Session $2$. The boxplots in Figure \ref{cigarettes-decrease-prozentual-boxplot} suggest that the  \textit{Lazy group} has a behaviour similar rather to the \textit{Motivated group} after six weeks but becoming different 
after six months. A similar observation can be derived from Figure \ref{cigarettes-bad-taste}.
Our interpretation is that the hypnosis sessions might have had some effects, but these effects were not sustainable for various reasons like an insufficient collaboration regarding the intervention protocol. In particular, we want to mention that several people in the \textit{Lazy group} consumed cigarettes within three hours after the first/second hypnosis sessions. Once again, this shall be regarded as an inadequate collaboration. In particular, those subjects did not listen to the audio files often, and therefore are assigned to the \textit{Lazy group}.
\par
The effect of the intensity of the individual post-session treatment (i.e., the number of audio listenings) is demonstrated in Figure \ref{fig:success-listenings} which suggests that participants shall be well-prepared in advance that they have to contribute to the success after the hypnosis sessions.
\par
Concerning the self-confidence and strength of wish in becoming non-smoker, the  \textit{Motivated group} had more self-confidence, but the \textit{Lazy group} had in average a higher strength of wish, see Subsection \ref{subsec:self-confidence}. The higher strength of wish in the \textit{Lazy group}  could possibly serve as an indicator for too high expectations. The lower self-confidence after Session $1$ in the \textit{Lazy group} can be interpreted as a hint that people have lost some trust in the success of  the intervention process, which could be the starting point for the lack of motivation afterwards. 
\par
Our conclusion is that people shall be motivated after Session $1$ to follow the intervention protocol and not to expect miracles; instead, they shall be aware of an ongoing intervention process which affords collaboration.
\par
As a summary, if a subject attended the hypnosis sessions in a serious manner and followed the intervention protocol as requested, then there is  a $86\%$ chance that this subject is \textit{abstinent of tobacco consumption after six months}.

\subsection{General Discussion}

Let us discuss the general setting of online hypnosis itself. When people visit a hypnotherapist in his office rooms, people can become nervous which maybe omits a sufficient deep state of trance. Vice versa, during an online hypnosis session people can stay at their comfort zone at home and can feel more safety due to the physical distance allowing better relaxation. However, according to our experience this distance can lead to much less seriousness in attending the hypnosis sessions: people may move around and stay not concentrated. A behaviour like this seems to be much less probable when people visit a hypnotherapist in person and pay a multiple for his service. This seems to lead to a rather higher drop-out rate for online hypnosis sessions compared to sessions in person. We have observed these effects with some people which were not considered for the evaluation of the study.
\par
We underline that this study is only a first evaluation of the efficiency of online hypnosis for smoking cessation. In particular, we neither made any comparison to alternative methods for smoking cessation nor we designed different groups of subjects a-priori. Furthermore, it was not possible to adapt the post-hypnotic suggestions for each single participant, which could possibly reduce the fall back rates after Session $1$ (e.g., some participants started to smoke the morning after). Therefore, it is strongly recommended to adapt the post-hypnotic suggestions individually, since every smoker has specific preferred situation in which he smokes. Post-hypnotic suggestions targeting these situations will very likely have an even better impact on the success probability. It remains open at this point how specific suggestions will improve the success rates.
Further open questions consider possible existence of \textit{a-priori} measureable, individual factors which influence the success rate.
\par
As a summary, we have seen that online hypnosis works well for smoking cessation under some premises. People have to be aware how hypnosis works, they shall participate in an appropriate manner (which means, they shall allow themselves to relax) and, in particular, they shall understand that hypnosis still needs some own motivation and some own effort (i.e., individual post-session treatment) for being successful. In other words, we may conclude: if a subject shows this discipline then the failure rate is quite low. At this point we clearly recommend that subjects are getting well-prepared before the start of the intervention and  subjects are motivated even after Session $1$ to follow the intervention protocol.

\section{Conflict of Interest}

Alexandra Wojak works as hypnocoach since 2019. Moreover, both authors founded a partnership under Civil Code (GbR) for hypnotical services in August 2020.

\section*{Acknowledgement}

The authors are grateful to the many participants for their collaboration. We also thank Lydia Heuser and Bernhard R\"ossler for their help to find participants.

\bibliographystyle{abbrv}
\bibliography{literature}

\end{document}